\documentstyle[stwol]{article}

\def\Journal#1#2#3#4{{#1} {\bf #2}, #3 (#4)}

\def\NPB{{\em Nucl. Phys.} B}
\def\PLB{{\em Phys. Lett.}  B}
\def\PRL{\em Phys. Rev. Lett.}
\def\PRD{{\em Phys. Rev.} D}

\def\be{\begin{equation}}
\def\ee{\end{equation}}
\def\bea{\begin{eqnarray}}
\def\eea{\end{eqnarray}}

\begin{document}

\title{NEW RESULTS ON TOPOLOGICAL SUSCEPTIBILITY IN SU(3) 
GAUGE THEORY}

\author{ B. ALLES, G. BOYD, M. D'ELIA, A. DI GIACOMO }

\address{Dipartimento di Fisica dell'Universit\`a and
INFN, Piazza Torricelli 2, 56126--Pisa, Italy}


\twocolumn[\maketitle\abstracts{We survey recent lattice results on 
QCD topological properties. The behaviour of the 
topological susceptibility at the deconfining phase transition
has been determined. This advance has been made possible by
an {\it i)} an improvement of the topological charge operator and
{\it ii)} a non-perturbative determination of renormalizations.}]

\section{Introduction}

Relevant progress in the study of topological properties of QCD has been 
made possible by two recent developments in lattice gauge theories:\\
1) Improvement of the operators \cite{chr}.\\
2) Non perturbative determination of the renormalizations \cite{vic}.

To clarify the meaning of 1) we recall that the building block of 
lattice gauge theories is the link, $U_{\mu}(\vec{n}) = 
\exp({i a A_{\mu}(\vec{n})})$, or parallel transport from the site
$\vec{n}$ of discretized spacetime to the neighbouring site in the direction 
$\hat{\mu}$, $\vec{n} + a \hat{\mu}$ ($a$ is the lattice spacing). All the 
operators
are constructed in terms of links, and therefore contain arbitrarily 
high powers of $a \,A_{\mu}$. The identification with continuum is usually 
done by requiring that the leading order in the power expansion of the 
lattice operator $O_L$ concides with its continuum counter-part $O$
\begin{equation}
O_L = O a^d + {\cal O}(a^{d+1})
\label{contlim}
\end{equation}
$d$ is the dimension of the operator. Higher order terms in 
eq. (\ref{contlim}) can be changed with large arbitrariness.

For the action, for example, eq. (\ref{contlim}) reads
\begin{equation}
S_L = -{1 \over 4} G_{\mu \nu} G^{\mu \nu} a^4 + {\cal O}(a^6)
\label{act-constr}
\end{equation}
Wilson's action $S_L^W = \beta \left(1 - 1/3\; {\rm Re}\; {\rm Tr} \Box
\right) $ obeys eq. (\ref{act-constr}).

The idea of improvement consists in exploiting the 
arbitrariness in higher order terms to reduce lattice artifacts. 
Improving the action can make the lattice larger in physical 
units \cite{lep}. Our approach will be to keep the usual Wilson action
and to improve the operator $Q_L$ for the topological charge~\cite{chr}, 
to reduce renormalizations from lattice to continuum. 
Renormalizations will be determined non perturbatively \cite{vic,all,all2}.

\section{Topology in QCD.}

The key relation is the anomaly of the singlet axial current
\be
\partial^{\mu} j^5_{\mu} = 2 N_f Q(x)
\label{anomaly}
\ee
where 
\be
j^5_{\mu} = \sum_f \bar{\psi_f} \gamma_{\mu} \gamma^5 \psi_f
\ee
$f$ means flavour, and $N_f$ number of flavours. $Q(x)$ is the topological
charge density
\be
Q(x) = {{g^2} \over {64 \pi^2}} \epsilon^{\mu \nu \rho \sigma} 
 G^a_{\mu \nu} (x) G^a_{\rho \sigma} (x).
\ee

The anomaly can explain the magnitude of the $\eta '$ mass (the so called 
$U(1)$ problem \cite{wei}) if the topological 
susceptibility of the vacuum in the 
quenched approximation (leading order in $1 \over {N_c}$ expansion)
\begin{equation}
 \chi \equiv \int d^4 x \langle 0 | T(Q(x) Q(0))| 0 \rangle. 
\label{deftopsusc}
\end{equation}
is large enough. Quantitatively \cite{wit,ven}
\begin{equation}
  {{2 N_f} \over {f_{\pi}^2}} \chi = m_{\eta }^2 + m_{\eta '}^2 
                       - 2 m_{K}^2.
\label{eq:massform}
\end{equation}

Testing eq. (\ref{eq:massform}) is a check at the same time of QCD 
and of the $1 \over {N_c}$ expansion, and can be done on the lattice.

The behaviour of $\chi$ at the deconfining temperature is also an important
test of models of QCD vacuum \cite{shu}. 

Another use of eq. (\ref{anomaly}) which can be made on the lattice is 
the measurement of the singlet axial charge of the nucleon, $G_1 (0)$, 
which is given by \cite{man}
\be
G_1 (0) \simeq \lim_{\vec{p} \rightarrow \vec{p} '} {N_f \over M_p}
{{\langle \vec{p} ' s ' | Q | \vec{p}  s \rangle} \over
{\bar{u}_{s '} (\vec{p} ') \gamma^5 u_s (\vec{p})}}
\ee
$G_1$ is usually related to the quark spin content of the proton \cite{ell}.

\section{Topology on the lattice.}

A lattice version of the topological charge density operator is
\begin{equation}
Q_L(x) = {{-1} \over {2^9 \pi^2}} 
\sum_{\mu\nu\rho\sigma = \pm 1}^{\pm 4} 
{\tilde{\epsilon}}_{\mu\nu\rho\sigma} \hbox{Tr} \left( 
\Pi_{\mu\nu}(x) \Pi_{\rho\sigma}(x) \right).
\label{stcharge}
\end{equation}
In the formal limit $a \rightarrow 0$, in accordance with eq. (\ref{contlim})
\begin{equation}
Q_L(x) = a^4 \, Q(x) + {\cal O}(a^{6}).
\label{ql}
\end{equation}
Any other choice for $Q_L$ will differ from (\ref{stcharge}) by 
${\cal O}(a^{6})$.
 
$Q_L(x)$ renormalizes multiplicatively
\be 
Q_L(x) \simeq_{\beta \rightarrow \infty} Z(\beta) Q(x) a(\beta)^4 + 
{\cal O}(a^6)
\ee

Similarly, for the lattice topological susceptibility, defined as
\be
\chi_L \equiv \langle \sum_x Q_L(x) Q_L(0) \rangle,
\ee
we have~\cite{cam}
\begin{eqnarray}
\chi_L &=&_{\beta \to \infty} Z(\beta)^2 \, a(\beta)^4 \,\chi +
M(\beta) \\
M(\beta)&\equiv& B(\beta) \langle G_2 \rangle + P(\beta) \langle
{\bf 1} \rangle + {\cal O}(a^6) \nonumber
\label{toplattice}
\end{eqnarray}
where $G_2$ is the trace of the energy-momentum tensor and {\bf 1} is the
identity operator. The presence of $M(\beta)$ is due to the fact that
the lattice regularization does not obey the prescription for the 
singularity at $x = 0$ which defines $\chi$ 
of eq. (\ref{deftopsusc}, \ref{eq:massform}) \cite{wit,ven}.
 From eq.(\ref{toplattice}) 
\be
\chi = {{\chi_L - M(\beta)} \over {Z^2 a^4 }}.
\label{topcontinuum}
\ee

$\chi_L$ is determined numerically on the lattice. $P(\beta)$,
$Z(\beta)$ and $B(\beta)$ depend on the choice of $Q_L$, i.e. on the
specification of the terms ${\cal O}(a^6)$ in eq. (\ref{ql}). A good
choice gives a small $M(\beta)$ and $Z \approx 1$, so that in 
eq. (\ref{topcontinuum}) $\chi\approx\chi_L$ and lattice artifacts 
are a small part of the numerical determination.

To determine $Z$, $Q_L \equiv \int {\rm d}^4x Q_L(x) \equiv Z\, Q$
is measured on an ensemble of configurations 
with a definite value of $Q$. A one instanton
configuration is heated to dress it with short range quantum fluctuations.
The number of instantons is checked step by step~\cite{del}, 
and the configurations
in which either the initial instanton has disappeared or new instantons 
have been created are discarded from the sample (Fig. 1). 
$M(\beta)$ is determined by measuring $\chi_L$ on an ensemble of 
 configurations with
$Q=0$ (0 instantons plus quantum fluctuations). Here again the global 
topological charge is checked step by step and configurations 
in which it has been changed in the heating procedure are 
discarded (Fig. 2). We have used an
improved operator obtained from $Q_L(x)$ by successive smearings,
and which differs from it by terms ${\cal O}(a^{6})$ \cite{chr}.

Figure 3 shows the plateaux reached by heating 1 instanton and the
$Z$'s for the 0, 1 and 2 improved operators. 

\vskip 5mm

\begin{figure}[htb]
\vspace{4.5cm}
\includegraphics{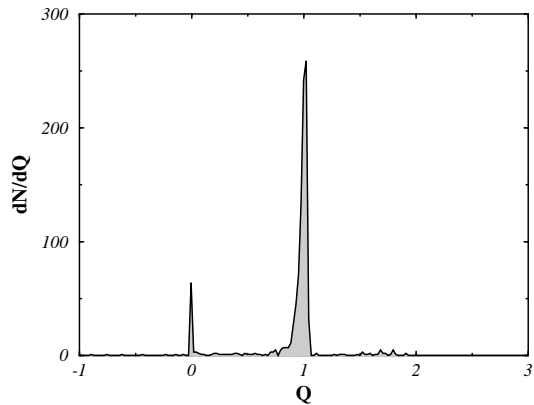} 
\null\vskip 0.3cm
\caption{Distribution of topological charge $Q$ for a set of 2000 
configurations obtained by 15 heat-bath updatings of a 1-instanton
configuration. $\beta = 5.75$.}
\end{figure}

\vskip 5mm

\begin{figure}[htb]
\vspace{4.5cm}
\includegraphics{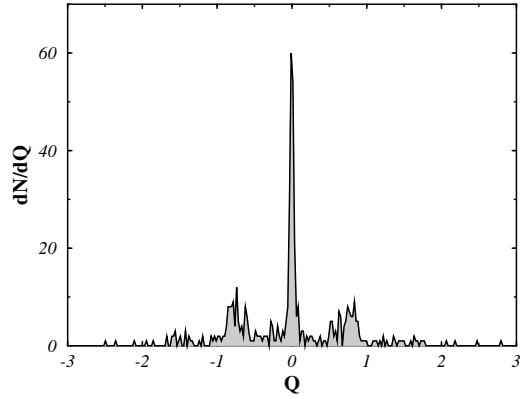} 
\null\vskip 0.3cm
\caption{Distribution of topological charge $Q$ for a set of 500 
configurations obtained by 36 heat-bath updatings of the zero field
configuration. $\beta = 5.90$.}
\end{figure}

\vskip 5mm

The analogous behaviour
for $M(\beta)$ is shown in Figure 4. Improving twice the operator 
produces a factor of 3 in $Z$, or a factor of 10 in $Z^2$, and a factor
of 10 reduction of $M(\beta)$. The ratio between the physical signal and
the lattice artifacts is then improved by 2 orders of magnitude going from
the 0 to the 2 smeared operator, and the subtraction in 
eq. (\ref{topcontinuum}) becomes of the order of 10\% of $\chi_L$ in
the scaling window.

\vskip 5mm

\begin{figure}[htb]
\vspace{4.5cm}
\includegraphics{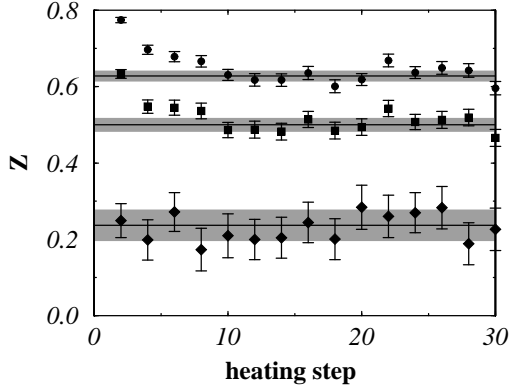} 
\null\vskip 0.3cm
\caption{Determination of $Z$ by heating of a 1 instanton configuration.
Diamonds, squares and circles correspond to 0, 1 and 2 smearings 
respectively. $\beta=6.36$.}
\end{figure}

\vskip 5mm

\begin{figure}[htb]
\vspace{4.5cm}
\includegraphics{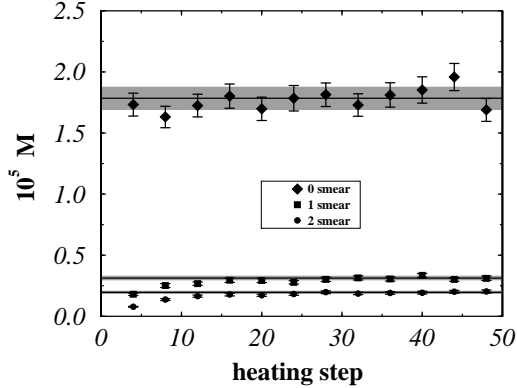} 
\null\vskip 0.3cm
\caption{Determination of $M(\beta)$ by heating of the flat configuration.
Diamonds, squares and circles correspond to 0, 1 and 2 smearings 
respectively. $\beta=6.30$.}
\end{figure}

\vskip 5mm

Figure 5 shows the result for $\chi$ at $T=0$ determined with the
0, 1 and 2 smeared operators: they agree with each other, scale properly
and confirm previous results in favour of the Witten-Veneziano mechanism
\cite{all,cam}. 

\vskip 5mm

\begin{figure}[htb]
\vspace{4.5cm}
\includegraphics{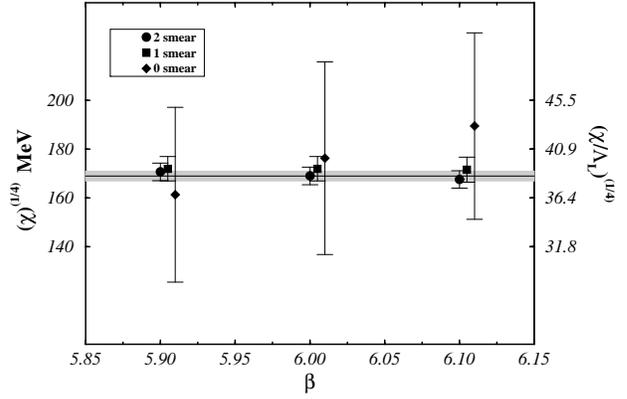} 
\null\vskip 0.3cm
\caption{Determination of $\chi$ at $T=0$. Diamonds, squares and
circles correspond to 0, 1 and 2 smearings respectively.}
\end{figure}

\vskip 5mm

Figure 6 shows a new result, which has been made possible by the improvement:
the behaviour of $\chi$ across $T_c$. The $\chi$ drops by at least
one order of magnitude at $T_c$. This result can provide an important 
check of models of QCD vacuum \cite{shu}.

\vskip 5mm

\begin{figure}[htb]
\vspace{4.5cm}
\includegraphics{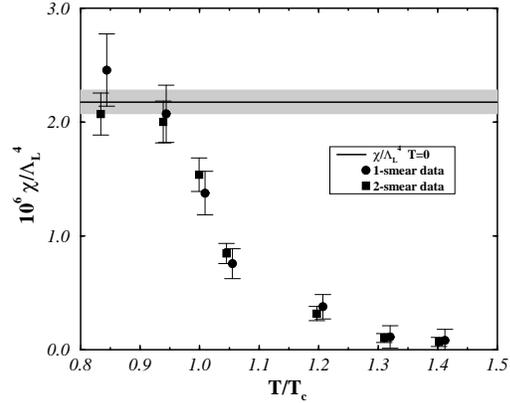} 
\null\vskip 0.3cm
\caption{Determination of $\chi$ at $T\not= 0$. Circles and squares
correspond to 1 and 2 smearings respectively. The horizontal line is the 
$T=0$ result.}
\end{figure}

\vskip 5mm

\noindent Difficulties are encountered in
thermalizing the topological charge with the hybrid Monte Carlo 
algorithm in full QCD \cite{boy} (Figs. 7,8,9): we are, however, implementing 
a multicanonical algorithm in which the quark mass is used as a new 
variable in the Monte Carlo, to exploit the fact that for high quark
masses this difficulty is less severe. The procedure works,  so that
the same technique described here will soon allow the determination of 
$\chi$, $\chi '$ in full QCD and the measurement of the spin content
of the proton.

\vskip 5mm

\begin{figure}[htb]
\vspace{4.5cm}
\includegraphics{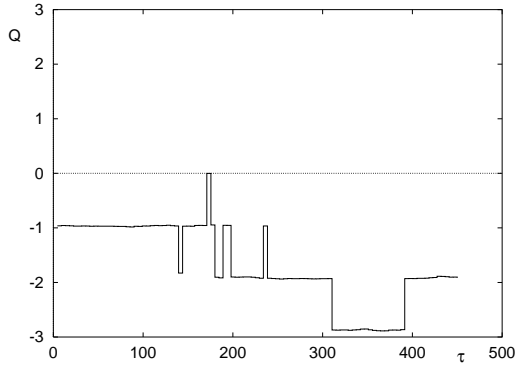} 
\null\vskip 0.3cm
\caption{
  Time history, in units of molecular dynamics time, of the topological
  charge $Q$ for a hybrid Monte Carlo simulation at $\beta=5.35$ and 
$a\,m=0.01$.}
\end{figure}

\vskip 5mm

\begin{figure}[htb]
\vspace{4.5cm}
\includegraphics{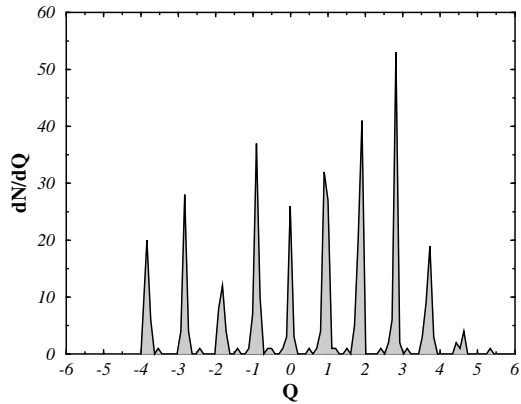} 
\null\vskip 0.3cm
\caption{Distribution of values of $Q$ in a sample of 1000 configurations
in full QCD obtained with the hybrid Monte Carlo algorithm.
$\langle Q \rangle \not= 0$ and the distribution is not symmetric under
$Q \to -Q$ in contrast with what happens with usual heat bath algorithms
for quenched QCD, see Fig. 9.}
\end{figure}

\vskip 5mm

\begin{figure}[htb]
\vspace{4.5cm}
\includegraphics{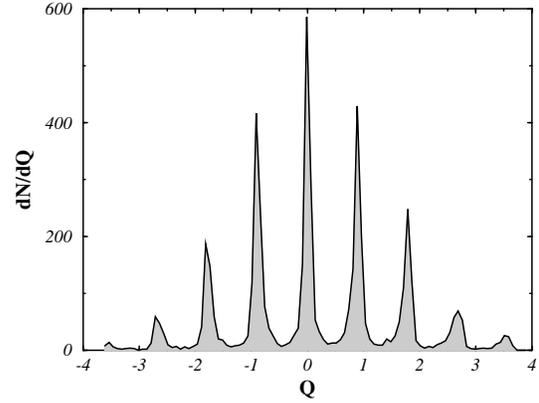} 
\null\vskip 0.3cm
\caption{Distribution of values of $Q$ 
in quenched QCD obtained with the heat bath algorithm.}
\end{figure}


\vskip 5mm


\section*{References}
\begin{thebibliography}{99}
\bibitem{chr} C. Christou, A. Di Giacomo, H. Panagopoulos, E. Vicari,
\Journal{\PRD}{53}{2619}{1996}.

\bibitem{vic} A. Di Giacomo,  E. Vicari, \Journal{\PLB}{275}{429}{1992}.

\bibitem{lep} G. P. Lepage, \Journal{\em Nucl. Phys. B (Proc. Suppl.)}{47}{3}{1996}.

\bibitem{all} B. All\'es, M. Campostrini, A, Di Giacomo, Y.
 G\"und\"u\c c, E. Vicari, \Journal{\PRD}{48}{2284}{1993}  

\bibitem{all2} B. All\'es, M. Campostrini, A, Di Giacomo, Y.
 G\"und\"u\c c, E. Vicari,  {\em Nucl. Phys. B (Proc. Suppl.)}
{\bf 34} 504 (1994).

\bibitem{wei} S. Weinberg, \Journal{\PRD}{11}{3583}{1975}.

\bibitem{wit} E. Witten, \Journal{\NPB}{156}{269}{1979}.

\bibitem{ven} G. Veneziano, \Journal{\NPB}{159}{213}{1979}.

\bibitem{shu} E. Shuryak, \Journal{\em Comments in Nuclear and Particle
Physics}{21}{235}{1994}. 

\bibitem{man} J. E. Mandula, \Journal{\PRL}{65}{1403}{1990}.

\bibitem{ell} J. Ellis, I. Karliner, \Journal{\PLB}{313}{213}{1993}.

\bibitem{cam} M. Campostrini, A. Di Giacomo, H. Panagopoulos, E. Vicari,
\Journal{\NPB}{329}{683}{1990}.

\bibitem{del} B. All\'es, M. D'Elia, A. Di Giacomo,  
hep-lat/9605013.  


\bibitem{boy} B. All\'es, G. Boyd, M. D'Elia, A. Di Giacomo, E. Vicari, 
hep-lat/9607049, to appear in {\em Phys. Lett.} B.

\end{thebibliography}
\end{document}